\documentclass{article} 
\usepackage{nips13submit_e,times}
\usepackage{hyperref}
\usepackage{url}

\usepackage[pdftex]{graphicx}
\usepackage{caption}
\usepackage{subcaption}

\let\a=\alpha     \let\d=\delta

 \let\D=\Delta   
         
  \let\r=\rho

\def\MM{{\cal M}}

\def\ttG{\tilde{G}}

\def\to{\rightarrow}

\newcommand{\beq}{\begin{equation}}
\newcommand{\eeq}{\end{equation}}

\title{Blind Calibration in
  Compressed Sensing using Message Passing Algorithms}

\author{
Christophe Sch\"ulke \\
ESPCI and CNRS UMR 7083\\
10 rue Vauquelin,\\
Paris 75005, France\\
\And
Francesco Caltagirone \\
Institut de Physique Th\'eorique\\
CEA Saclay and CNRS URA 2306\\
 91191 Gif-sur-Yvette, France \\
\AND
Florent Krzakala \\
ESPCI and CNRS UMR 7083\\
10 rue Vauquelin,\\
Paris 75005, France\\
\And
Lenka   Zdeborov\'a \\
Institut de Physique Th\'eorique\\
CEA Saclay and CNRS URA 2306\\
 91191 Gif-sur-Yvette, France \\
}

\nipsfinalcopy 

\begin{document}

\maketitle

\begin{abstract} 
Compressed sensing (CS) is a concept that allows to acquire compressible signals with a
small number of measurements. As such it is very attractive for
hardware implementations. Therefore, correct calibration of the hardware is 
a central issue. In this paper we study the so-called blind
calibration, i.e. when the training signals that are available to perform the 
calibration are sparse but unknown. We
extend the approximate message passing (AMP) algorithm used in CS to the case of blind
calibration. In the calibration-AMP, both the gains on the
sensors and the elements of the signals are treated as unknowns. 
Our algorithm is also applicable to  settings in which the sensors 
distort the measurements in other ways than multiplication by a gain, unlike previously suggested blind calibration
algorithms based on convex relaxations.  We study numerically the phase
diagram of the blind calibration problem, and show that even in
cases where convex relaxation is possible, our algorithm requires a
smaller number of measurements and/or signals in order to perform
well. 
\end{abstract}

\section{Introduction}

The problem of acquiring an $N$-dimensional signal ${\bf x}$ through $M$ linear
measurements, ${\bf y}=F{\bf x}$, arises in many contexts. 
The Compressed Sensing (CS) approach \cite{CandesTao:05,Donoho:06}
exploits the fact that, in many cases of interest, the signal is $K$-sparse (in an
appropriate known basis), meaning that only $K=\r N$ out of the $N$
components are non-zero. Compressed sensing theory shows that a $K$-sparse
$N$-dimensional signal can be reconstructed from far less than $N$
linear measurements
\cite{CandesTao:05,Donoho:06}, thus saving acquisition
time, cost or increasing the resolution. In the most common setting, the linear
$M \times N$ map $F$ is considered to be known. 

Nowadays, the concept of compressed
sensing is very attractive for hardware implementations. However, one of the
main issues when building hardware revolves around calibration. 
Usually the sensors introduce a distortion (or decalibration) to the measurements in the
form of some unknown gains. Calibration is about how to determine the transfer function
between the measurements and the readings from the sensor. In some applications
 dealing with distributed sensors or
radars for instance, the location or intrinsic parameters of the sensors are not
exactly known \cite{NgSee96,YangZhang12}. Similar distortion can be found in applications with microphone arrays \cite{MignotDaudet11}.
The need for calibration has been emphasized in a number of other works, see
e.g. \cite{RaghebLaska08,TroppLaska10,PankiewiczArildsen13}.  One common way of dealing with calibration (apart from ignoring it or
considering it as measurement noise) is {\it supervised
calibration} when some known training signals ${\bf x}_l$, $l=1,\dots,P$ and the corresponding
observations ${\bf y}_l$ are used to estimate the distortion
parameters. 

In the present work we are interested in {\it blind (unsupervised) calibration}, in which
{\it known} training signals are not available, and one can only use
{unknown} (but sparse) signals. If such calibration
 is computationally possible, then it might be
simpler to do than the {\it supervised calibration} in practice .

\subsection{Setting}

We state the problem of {\it blind calibration} in the following way.
First we introduce an unknown distortion parameter (we will also use
equivalently the term decalibration parameter or gain) $d_\mu$ for each of the
sensors, $\mu=1,\dots,M$. Note that $d_{\mu}$ can also represent a
vector of several parameters. We consider that the signal is linearly projected by
a known $M\times N$ measurement matrix $F$ and only then distorted
according to some known transfer function $h$. This transfer function
can be probabilistic (noisy), non-linear, etc. Each sensor $\mu$ then
provides the following distorted and noisy reading (measure)
$y_{\mu}=h(z_{\mu l},d_{\mu},w_{\mu})$ where
$z_{\mu}=\sum_{i=1}^N F_{\mu i} x_{i}$ is the linear
projection of the signal on the $\mu$th row of the measurement matrix
$F$. For the measurement noise $w_{\mu}$, one usually one considers an
 iid Gaussian noise with variance $\Delta$ added to $z_{\mu}$.

In order to perform the blind calibration, we need to measure several
statistically diverse signals. 
Given a set of $N$-dimensional $K$-sparse signals ${\bf x}_l$ with
$l=1, \cdots , P$, for each of the signals we consider $M$ sensor readings 
\begin{equation}
y_{\mu l} = h(z_{\mu l},d_{\mu},w_{\mu l}) \, ,  \quad {\rm where} \quad  z_{\mu l} = \sum_{i=1}^N F_{\mu i} x_{il} \, ,\label{definition}
\end{equation}
where $d_\mu$ are the signal-independent distortion parameters, $w_{\mu
l}$ is a signal-dependent measurement noise, and $h$ is an arbitrary
known function of these variables with standard regularity
requirements. Given the $M \times P$ measurements $y_{\mu l}$ and a perfect knowledge of the matrix $F$, we 
 want to infer both the $P$ different signals $\lbrace {\bf x}_1, \cdots {\bf x}_P \rbrace$ and the $M$ distortion parameters $d_{\mu}$, $\mu=1, \cdots M$. 

\subsection{Relation to previous work}


As far as we know, the problem of blind calibration was first studied in the
context of compressed sensing in \cite{GribonvalChardon11} where the distortions were
considered as multiplicative, i.e. the transfer function was
\begin{equation}
      h(z_{\mu l},d_{\mu},w_{\mu l})   = \frac{1}{d_\mu} ( z_{\mu l}+
      w_{\mu l}) \, .    \label{product}
\end{equation}
A subsequent work \cite{BilenPuy13} considers a more general case when
the distortion parameters are $d_\mu = (g_\mu,\theta_\mu)$, and the
transfer function  $h(z_{\mu l},d_{\mu},w_{\mu l})   = e^{i \theta_\mu} ( z_{\mu l}+
      w_{\mu l})/g_\mu $. Both \cite{GribonvalChardon11} and  \cite{BilenPuy13}  applied convex
optimization based algorithms to the blind calibration problem and their
approach seems to be limited to the above special cases of transfer
functions. Our approach is able to
deal with a general transfer function $h$, and moreover for the
product-transfer-function (\ref{product}) it outperforms the algorithm
of \cite{GribonvalChardon11}. 

The most commonly used algorithm for signal reconstruction in
compressed sensing is the $\ell_1$ minimization of
\cite{CandesTao:05}. In compressed sensing without
noise and for measurement matrices with iid Gaussian elements, the
$\ell_1$ minimization algorithm leads to exact reconstruction as long
as the measurement rate  $\alpha = M/N > \alpha_{\rm DT}$ in the limit
of large signal dimension, where $\alpha_{\rm DT}$ is a well known phase transition of
Donoho and Tanner \cite{Donoho05072005}. The blind calibration
algorithm of  \cite{GribonvalChardon11,BilenPuy13} also directly uses
 $\ell_1$ minimization for reconstruction.

In the last couple of years, the theory of compressed sensing witnessed
a large progress thanks to the development of message
passing algorithms based on the standard loopy Belief Propagation (BP)
and their analysis \cite{DonohoMaleki09,DonohoMaleki10,Rangan10b,KrzakalaPRX2012,DonohoJavanmard11}.  
In the context of compressed sensing, the canonical loopy BP is difficult to implement
because its messages would be probability distributions over a
continuous support. At the same time in problems such as compressed
sensing, Gaussian or quadratic approximation of BP still contains the
information necessary for a successful reconstruction of the
signal. Such approximations of loopy BP originated in works on CDMA
multiuser detection \cite{BoutrosCaire02,Kabashima03}. 
In compressed
sensing the Gaussian approximation of BP is known as the approximate
message passing (AMP)
\cite{DonohoMaleki09,DonohoMaleki10}, and it was used to prove that
with properly designed measurement matrices $F$ the signal can be
reconstructed for measurement rate as low as $\alpha= \rho$
asymptotically, thus closing the gap between the Donoho-Tanner
transition and the information theoretical lower bound \cite{KrzakalaPRX2012,DonohoJavanmard11}. Even without
particular design of the measurement matrices the AMP algorithm
outperforms the $\ell_1$-minimization for a large class of
signals. Importantly for the present work, \cite{Rangan10b} generalized the AMP
algorithm to deal with a wider range of input and
output functions. For some of those, generalizations of the $\ell_1$-minimization based
approach are not convex anymore, and hence they do not have the
advantage of provable computational tractability anymore.


The following two works have considered blind calibration related problems
with the use if AMP-like algorithms. In \cite{KamilovBourquard2013}
the authors use AMP combined with expectation maximization to
calibrate gains that act on the signal components rather than on the measurement
components as we consider here. In \cite{KrzakalaMezard13} the
authors study the case when every element of the measurement matrix
$F$ has to be calibrated, in contrast to the row-constant gains
considered in this paper. The setting of \cite{KrzakalaMezard13}  is
much closer to the dictionary learning problem and is much more
demanding, both computationally and in terms of the number of different
signals
necessary for successful calibration.

\subsection{Contributions}


In this work we extend the generalized approximate message passing (GAMP)
algorithm of \cite{Rangan10b} to the problem of blind calibration with
a general transfer function $h$, eq.~(\ref{definition}). We denote it
as the calibration-AMP or C-AMP algorithm. The C-AMP uses
$P>1$ unknown sparse signals to learn both the different signals ${\bf
x}_l$, $l=1,\dots,P$, and the
distortion parameters $d_\mu$, $\mu =1, \dots, M$, of the sensors. We hence overcome the limitations of the blind calibration algorithm
presented in \cite{GribonvalChardon11,BilenPuy13} to the class
of settings for which the calibration can be written as a convex optimization problem. 

In the second part of this paper we analyze the performance of C-AMP for the product transfer function (\ref{product}) used in
\cite{GribonvalChardon11} and demonstrate its scalability and
better performance with respect to their $\ell_1$-based approach. In the numerical study we observe a sharp phase transition generalizing
the phase transition seen for AMP in compressed sensing
\cite{KrzakalaMezard12}. Note that for the blind calibration problem to be solvable, we need the
amount of information contained in the sensor readings,  $PM$,  to be at least as large
as the size of the vector of distortion parameters $M$, plus the number of 
the non-zero components of all the signals, $KP$. Defining 
$\r=K/N$ and $\a=M/N$, this leads to $\a  P \geq \r  P + \a$.
If we fix the number of signals $P$ we have a well defined line in the $(\r, \a )$-plane given by 
\beq
\a \geq \frac{P}{P-1} \r \equiv \alpha_{\rm min} \, ,
\label{counting}
\eeq
below which exact calibration cannot be possible. 
We will compare the empirically observed phase transition to this
theoretical bound as well as to the phase transition that would have
been observed in the pure compressed sensing, i.e. if we knew the distortion parameters.

\section{The Calibration-AMP algorithm}
The approximate message passing algorithm is based on a Bayesian
probabilistic formulation of the reconstruction problem. Denoting
$P_X(x_{il})$ the assumed empirical distribution of the components of the
signal, $P_W(w_{\mu l})$ the assumed probability distribution of the
components of the noise, and $P_D(d_\mu)$ the assumed empirical distribution of
the distortion parameters, the Bayes formula yields
\beq
\label{bayes}
P({\bf x},{\bf d}|{\bf F},{\bf y})= \frac{1}{Z}\prod_{i,l=1}^{N,P}
P_{X}(x_{il})\, \prod_{\mu=1}^{M} P_{D}(d_\mu) \prod_{l,\mu=1}^{P,M}
\int \, dw_{\mu l} P_W(w_{\mu l}) \d \left[ y_{\mu l} - h\left(z_{\mu
      l} ,d_{\mu}, w_{\mu l}\right) \right]\, , \eeq where $Z$ is a
normalization constant and $z_{\mu l} = \sum_i F_{\mu i}
x_{il}$. We denote the marginals of the signal components $\nu^x_{il}(x_{il})=\int \prod_{\mu} {\rm
  d}d_{\mu} \prod_{jn\neq il} dx_{jn} \,P({\bf x},{\bf d}|{\bf F},{\bf
  y})$ and those of  the distortion parameters $\nu^x_{il}(x_{il})=\int \prod_{\mu} {\rm
  d}d_{\mu} \prod_{jn\neq il} dx_{jn} \,P({\bf x},{\bf d}|{\bf F},{\bf
  y})$.  The estimators $x^*_{il}$ that minimizes the expected mean-squared
error (MSE) of the signals and the estimator $d^*_{\mu}$
of the distortion parameters are the averages w.r.t. the marginal
distributions, namely $x^*_{il}=\int {\rm d}x_{il}\, x_{il} \,
\nu^x_{il}(x_{il})$ and $d^*_{\mu}=\int {\rm d}d_{\mu} \, d_{\mu}\,
\nu^d_{\mu}(d_{\mu})$.  An exact computation of these estimates is not
tractable in any known way so we use instead a belief-propagation
based approximation that has proven to be fast and efficient in the
compressed sensing problem
\cite{DonohoMaleki09,DonohoMaleki10,Rangan10b}.

\begin{figure}[!ht]
\centering
\includegraphics[width=6cm]{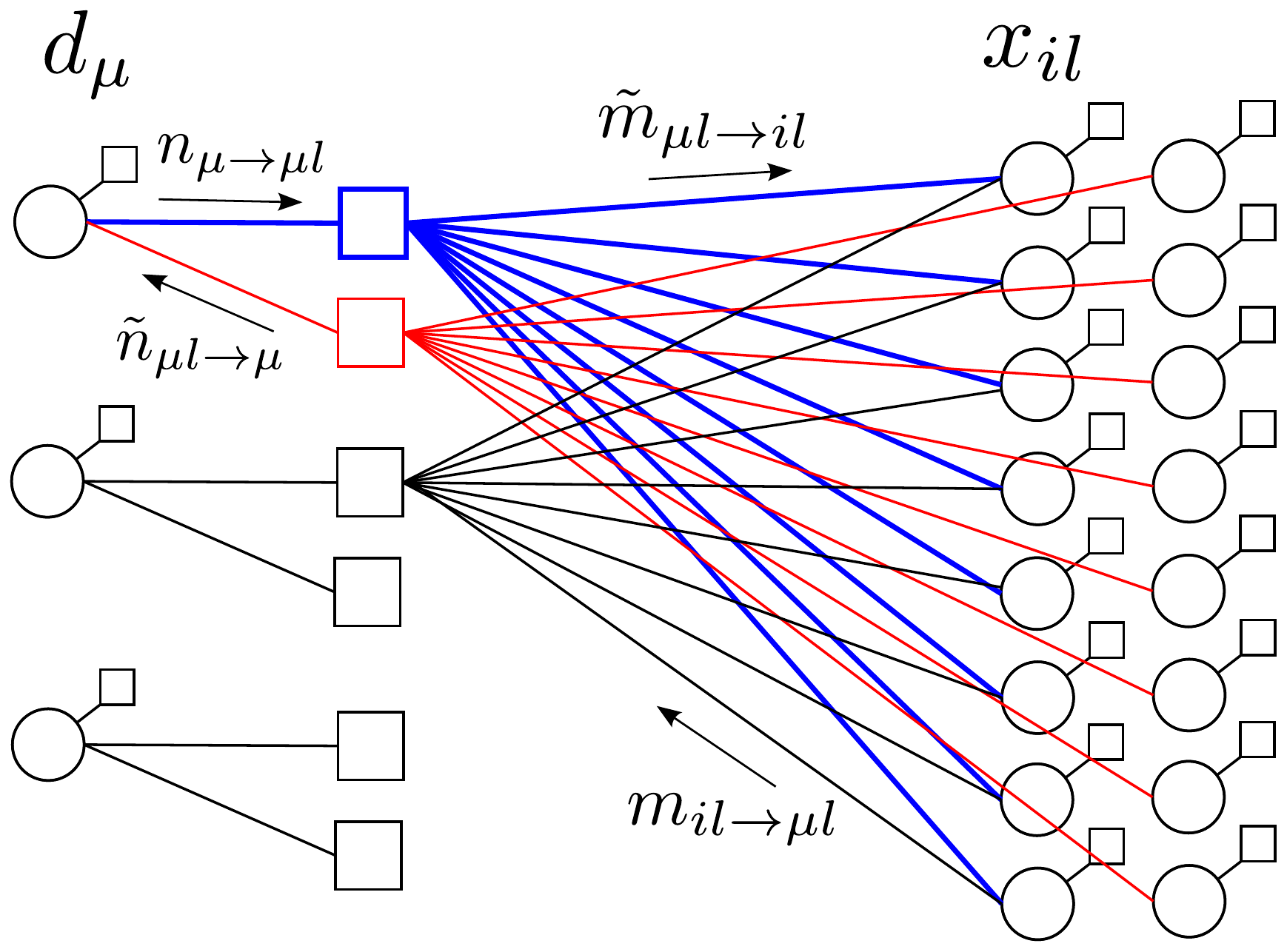}
\caption{\label{factro_graph} Graphical model representing the blind calibration
  problem.  Here the dimensionality of the signal is $N\!=\!8$, the
  number of sensors is $M\!=\!3$, and the number of signals used for
  calibration $P\!=\!2$. The variable nodes $x_{il}$ and $d_\mu$ are
  depicted as circles, the factor nodes as squares.}
\label{factorGraph}
\end{figure}

Given the factor graph representation of the calibration problem in Fig.~\ref{factorGraph},
the canonical belief propagation equations for the probability measure
(\ref{bayes}) are written in terms of $NPM$
messages $\tilde m_{\mu l \to il}(x_{il})$, and $m_{il\to \mu
  l}(x_{il})$ representing probability distributions on the signal component
$x_{il}$, and $PM$ messages $n_{\mu \to \mu l}$ and $\tilde n_{\mu l \to
  \mu}$ representing probability distributions on the distortion
parameter $d_\mu$. Following the lines of
\cite{DonohoMaleki09,DonohoMaleki10,Rangan10b,KrzakalaPRX2012}, with
the use of the central limit theorem, a Gaussian approximation, and
neglecting terms that go to zero as $N\to \infty$, the
BP equations can be closed using only the means $a_{il \to \mu l}=\int
{\rm d}x_{il} \, m_{il\to \mu l} (x_{il})\, x_{il}$ and variances
$v_{il \to \mu l}=\int {\rm d}x_{il} \, m_{il\to \mu l} (x_{il}) \, x^2_{il}-a^2_{il \to \mu l}$ of the
above messages $m_{il\to \mu l}$ and the means $k_{\mu \to \mu l}=\int
{\rm d}d_{\mu} \, n_{\mu\to \mu l} (d_{\mu})\, d_{\mu}$ and variances $l_{\mu \to \mu l}=\int
{\rm d}d_{\mu} \, n_{\mu\to \mu l} (d_{\mu})\, d^2_{\mu} - k^2_{\mu \to \mu l}$ of the
above messages $n_{\mu \to \mu l}$.  Moreover, again neglecting only terms that go to zero
as $N\to \infty$, we can write closed equations on quantities that
correspond to the variables and factors nodes, instead of messages
running between variables and factor nodes. For this we introduce  
$\omega_{\mu l}=\sum_i F_{\mu i} a_{il\to\mu l}$ and $V_{\mu l}=\sum_i
F^2_{\mu i} v_{il\to\mu l}$. The derivation of the C-AMP algorithm
goes very much along the lines of \cite{DonohoMaleki09,DonohoMaleki10,Rangan10b,KrzakalaPRX2012}.
To summarize the resulting algorithm we define the
``generating function'' $G$ as
\begin{equation}
G(y_{\mu \cdot},\omega_{\mu \cdot},V_{\mu \cdot}, \theta)=\ln
\left[\int {\rm d}d \, P_{D}(d) \, \prod_{n=1}^P \tilde{G} (y_{\mu n},
  d, \omega_{\mu n}, V_{\mu n}) \, e^{\theta d}\right]\, , \label{gen_fc}
\end{equation}
where
\begin{equation}
   \ttG(y,d,\omega, v)=\int  {\rm d}z \, {\rm d}w \, P_W(w)\,
   \d[h(z,d,w)-y] \, e^{-\frac{1}{2} \frac{(z-\omega)^2}{v}}\, ,
\end{equation}
where $\mu \cdot$ indicates a dependence on all the variables labeled
$\mu n$ with $n=1, \cdots, P$,  and $\d(\cdot)$ is the Dirac delta
function. 
Similarly as Rangan in \cite{Rangan10b}, we define $P$ output
functions as 
\beq
g^{l}_{\mathrm{out}}(y_{\mu \cdot},\omega_{\mu \cdot},V_{\mu \cdot})=
\frac{\partial}{\partial \omega_{\mu l}}G(y_{\mu \cdot},\omega_{\mu
  \cdot},V_{\mu \cdot},\theta=0)\, .
\eeq
Note that each of the output functions depend
on all the $P$ different signals. 
We also define the following input functions
\begin{equation}
f^x_a(\Sigma^2 , R)=[x]_{X} \, ,\quad \quad    f^x_c(\Sigma^2 ,
R)=[x^2]_{X}-[x]_{X}^2 \, , 
\end{equation}
where $[\dots]_{X}$ indicates expectation w.r.t. the measure 
\beq
      \MM_{X}(x,\Sigma^2 ,R) = \frac{1}{Z(\Sigma^2 ,R)} P_{X}(x)  \,
      \, e^{-\frac{(x-R)^2}{2\Sigma^2} }\, .
\eeq

Given the above definitions, the iterative calibration-AMP algorithm reads as follows 
\begin{eqnarray}
V^{t+1}_{\mu l}&=&\sum_i F_{\mu i}^2 \, v^{t}_{il}\, , \label{V_mul}\\
e^{t+1}_{\mu l}&=&g^{l}_{\mathrm{out}}(y_{\mu \cdot},\omega^{t}_{\mu
  \cdot},V^{t+1}_{\mu \cdot})\, ,\\
h^{t+1}_{\mu l}&=&-\frac{\partial}{\partial \omega_{\mu l}}
g^{l}_{\mathrm{out}}(y_{\mu \cdot},\omega^{t}_{\mu \cdot},V^{t+1}_{\mu
  \cdot})\, ,\\
\omega^{t+1}_{\mu l}&=&\sum_i F_{\mu i} a^{t}_{il} - V^{t+1}_{\mu l}
\, e^{t+1}_{\mu l}\, , \label{omega}\\
(\Sigma^{t+1}_{il})^2 =\left[ \sum_{\mu} F_{\mu i}^2 \, h^{t+1}_{\mu
    l} \right]^{-1} \, ,~&&~~~  R^{t+1}_{il}= a_{il} +  \left[ \sum_{\mu} F_{\mu i}\, e^{t+1}_{\mu
    l} \right] (\Sigma^{t+1}_{i l})^2\, ,\\
a^{t+1}_{il}=f^x_a((\Sigma^{t+1}_{il})^2,R^{t+1}_{il})\, , ~&&~~~ v^{t+1}_{il}=f^x_c((\Sigma^{t+1}_{il})^2,R^{t+1}_{il})\, ,  \label{v_il}
\end{eqnarray}
we initialize $\omega_{\mu l}^{t=0}=y_{\mu l}$, $a^{t=0}_{il}$ and
$v_{il}^{t=0}$ as the mean and variance of the assumed distribution
$P_X(\cdot)$, and iterate these equations until convergence. At every
time-step the quantity $a_{il}$ is the estimate for the signal element
$x_{il}$, and $v_{il}$ is the approximate error of this estimate.
The estimate and its error for the distortion parameter $d_{\mu}$ can
be computed as 
\begin{eqnarray}
k^{t+1}_{\mu}&=& \frac{\partial   }{ \partial \theta  } G(y^{t+1}_{\mu
  \cdot},\omega^{t+1}_{\mu \cdot},V^{t+1}_{\mu \cdot},\theta)
\Big{|}_{\theta = 0}\, ,\\
l^{t+1}_{\mu}&=&\frac{\partial ^2  }{ \partial \theta^2  }
G(y^{t+1}_{\mu \cdot},\omega^{t+1}_{\mu \cdot},V^{t+1}_{\mu
  \cdot},\theta) \Big{|}_{\theta = 0} \, .
\end{eqnarray}
By setting $P_D(d_{\mu})=\delta(d_\mu-d_\mu^{\rm true})$, and
simplifying eq.~(\ref{gen_fc}), readers familiar
with the work of Rangan \cite{Rangan10b} will recognize the GAMP
algorithm in eqs. (\ref{V_mul}-\ref{v_il}). Note that for a general
transfer function $h$ the generating function $G$ (\ref{gen_fc}) has to
be evaluated numerically. The overall running time of the C-AMP
algorithm is comparable to the one of GAMP \cite{Rangan10b}, it is
hence very scalable.

\subsection{C-AMP for the product transfer function}

In the numerical section of this paper we will focus on a specific
case of the transfer function $h(z_{\mu l},d_\mu,w_{\mu l})$, defined in
eq.~(\ref{product}). We consider the measurement noise $w_{\mu l}$ to
be Gaussian of zero mean and variance $\Delta$. This transfer function
was considered in the work of~\cite{GribonvalChardon11} and we will hence be able to
compare the performance of C-AMP directly to the convex optimization
investigated in \cite{GribonvalChardon11}. 
For the product transfer function eq.~(\ref{product}) most integrals
requiring a numerical computation in the general case are expressed
analytically and C-AMP becomes (together with eqs. (\ref{V_mul}) and
(\ref{omega}-\ref{v_il}):
\begin{eqnarray}
e^{t+1}_{\mu l}&=&\frac{k^t_{\mu} y_{\mu l} -\omega^t_{\mu l} }{
  V^{t+1}_{\mu l} + \D }\, ,\\
h^{t+1}_{\mu l}&=&\frac{1}{V^{t+1}_{\mu l} + \D} - \frac{l^{t}_{\mu}
  y_{\mu l}^2}{(V^{t+1}_{\mu l} + \D)^2}\, ,\\
k^{t+1}_{\mu}&=& f^d_a((C^{t+1}_{\mu})^2,T^{t+1}_{\mu})\, ,\\
l^{t+1}_{\mu}&=&f^d_c((C^{t+1}_{\mu})^2,T^{t+1}_{\mu}) \, ,\\
(C^{t+1}_{\mu})^2&=&\left[ \sum_{n}\frac{y_{\mu n}^2}{V^{t+1}_{\mu
      n}+\D} \right]^{-1}\, ,\\
T^{t+1}_{\mu}&=&(C^{t+1}_{\mu})^2 \, \sum_{n} \frac{y_{\mu n}
  \omega^{t+1}_{\mu n} }{V^{t+1}_{\mu n} + \D} \, . 
\end{eqnarray}
where we have introduced
\begin{equation}
f^d_a(C^2, T)=[d]_{D}\, , \quad \quad f^d_c(C^2, T)=[d^2]_{D}-[d]_{D}^2
\, ,
\label{d_update_functions}
\end{equation}
with $[\dots]_{D}$ indicating the expectation w.r.t. the measure 
\beq
\MM_{D}(d,C^2 ,T) = \frac{1}{Z(C^2 ,T)} P_{D}(d) |d|^P \,  \,
e^{-\frac{(d-T)^2}{2 C^2} } \, .
\eeq



\section{Experimental results}
Our simulations were performed using a MATLAB implementation of the
C-AMP algorithm presented in the previous section, that is available online. We focused on the
noiseless case $\Delta = 0$ for which, when there is no distortion of
the output, exact reconstruction of the signal is possible.
We tested the algorithm on randomly generated Gauss-Bernoulli signals
with density of non-zero elements $\rho$, their distribution being a
Gaussian one with zero mean and unit variance. For the present
experiments the algorithm is using this information via a matching
distribution $P_X(x_{il})$. The situation when $P_X$ mismatches the
true signal distribution was discussed for AMP for compressed sensing
in \cite{KrzakalaMezard12}.

The distortion parameters $d_\mu$ were generated from a uniform
distribution centered at $d=1$ having variance $\sigma^2$. This
ensures that, as $\sigma^2 \to 0$, the results of standard compressed
sensing are recovered, while the distortions are more and more serious
as $\sigma^2$ is growing.  For numerical stability purposes, the
variance of the assumed distribution of the distortions used in the
update functions of C-AMP was taken to be slightly larger than the
variance used to create the actual distortion parameters. For the same
reasons, we have also added a small noise $\Delta = 10^{-17}$ and used
a damping factor in the iterations in order to avoid oscillatory
behavior. In this noiseless case we iterate the C-AMP equations until
the following quantity ${\rm crit} = \frac{1}{MP} \sum_{\mu l} \left(
  k_{\mu} y_{\mu l} - \sum_i F_{\mu i} a_{il} \right)^2$ becomes
smaller than the numerical precision of implementation (in case of
perfect recovery), around $10^{-16}$, or until that quantity does not
decrease any more over 100 iterations (when a fixed point is reached,
but the reconstruction is not perfect).


Success or failure of the reconstruction is usually determined by looking at the mean squared error (MSE) between the true signal $\bf x_l^0$ 
and the reconstructed ones $\bf a_l$. 
In the noiseless setting the product transfer function $h$ leads to a scaling 
invariance: if $\bf x_l^0$ and $\bf d^0$ are the true signals and the
true distortion parameters, multiplying both by the same non-zero real number $s$ leads to another 
possible solution of the system.  Therefore, a better measure of success is the cross-correlation between real and 
recovered signal (used in \cite{BilenPuy13}) or a corrected version of the MSE, defined by:
\begin{equation}
 {\rm MSE}^{\rm corr} = \frac{1}{N P} \sum_{il} \left( x_{il}^0 -
   \hat{s} a_{il} \right)^2 \, , \quad {\rm where} \quad \hat{s}=
 \frac{1}{M} \sum_{\mu} \frac{d^0_{\mu}}{k_{\mu}} \label{MSE_corr}
\end{equation}
is an estimation of the scaling factor $s$.
Slight deviations between empirical and theoretical means due to the finite size of $M$ and $N$ lead to important 
differences between ${\rm MSE}$ and ${\rm MSE}^{\rm corr}$, only the
latter going truly to zero for finite $N$ and $M$.


\begin{figure}[!ht]
\begin{center}
\includegraphics[width=0.85\textwidth]{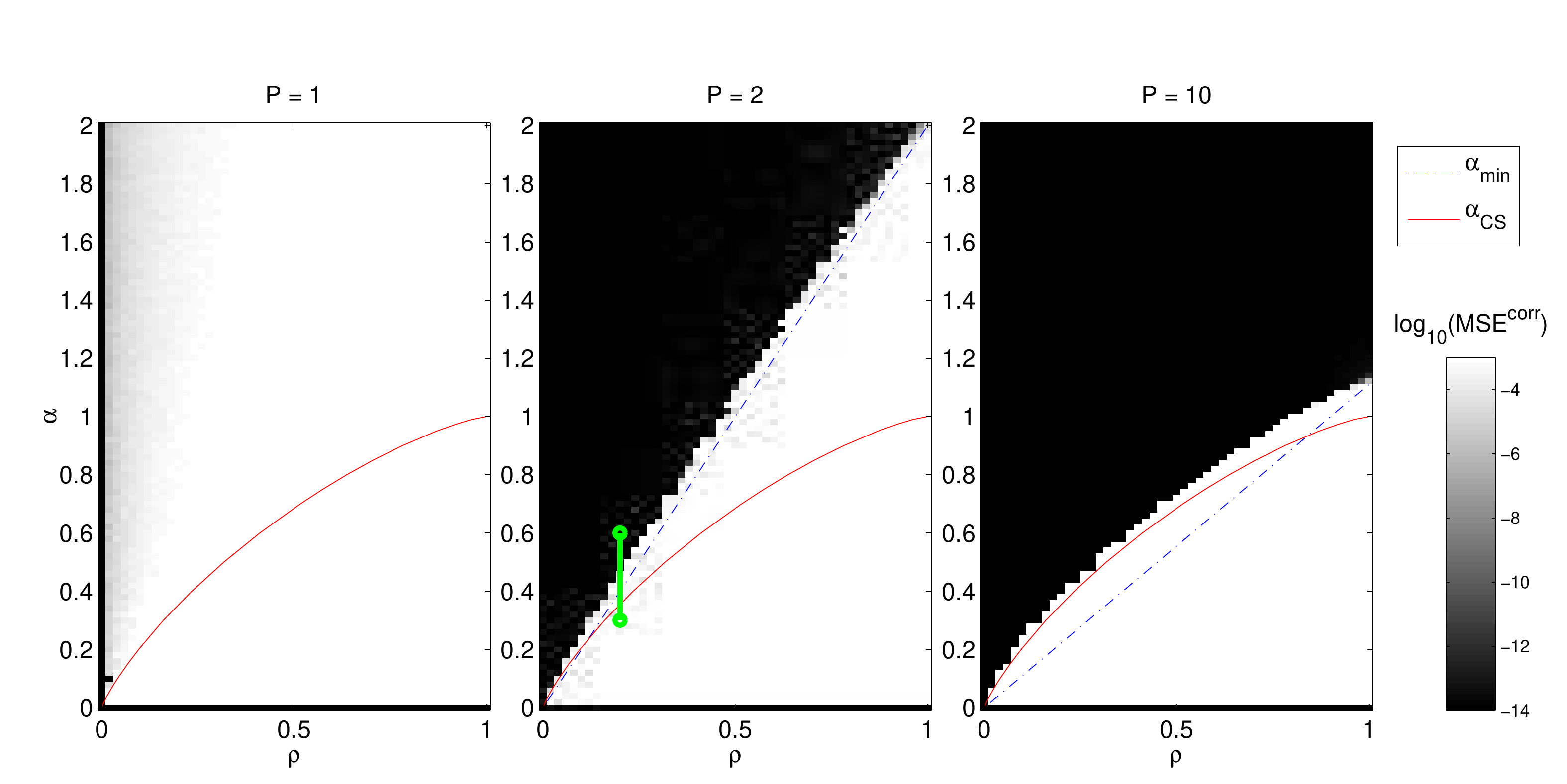}
\end{center}
\caption{Phase diagrams for different numbers $P$ of calibrating
  signals: The measurement rate $\alpha=M/N$ is plotted against the
  density of the signal $\rho=K/N$. The plotted value is the decimal
  logarithm of the corrected mean squared error (\ref{MSE_corr})
  achieved for one randomly chosen instance.  White indicates failure
  of the reconstruction, while black represents perfect reconstruction
  (i.e. a MSE of the order of the numerical precision). In this figure
  the distortion variance is $\sigma^2=0.01$ and $N=1000$. While for
  $P\!=\!1$ reconstruction is never possible, for $P\!>\!1$, there is
  a phase transition very close to the lower bound defined by
  $\alpha_{\rm min}$ in equation (\ref{counting}) or to the phase
  transition line of the pure compressed sensing problem $\alpha_{\rm
    CS}$. Note that while this diagram is usually plotted only for
  $\alpha\le 1$ for compressed sensing, the part $\alpha>1$ displays
  pertinent information in blind calibration.}
\label{Fig:ARdiags}
\end{figure}

Fig.~\ref{Fig:ARdiags} shows the empirical phase diagrams in the
$\alpha$-$\rho$ plane we obtained from the C-AMP algorithm for
different number of signals $P$.  For $P=1$ the reconstruction is
never exact, whereas for any $P>1$, there is a sharp phase transition
taking place with a jump in ${\rm MSE}^{\rm corr}$ of ten orders of
magnitude. As $P$ increases, the phase of exact reconstruction gets
bigger and tends to the one observed in Bayesian compressed sensing
when no distortion of the output is present \cite{KrzakalaPRX2012}.
Remarkably, for small values of the density $\rho$, the position of
the C-AMP phase transition is very close to the compressed sensing one
already for $P=2$ and C-AMP performs almost
as well as in the absence of distortion.

\begin{figure}[!ht]
\centering
\includegraphics[width=\textwidth]{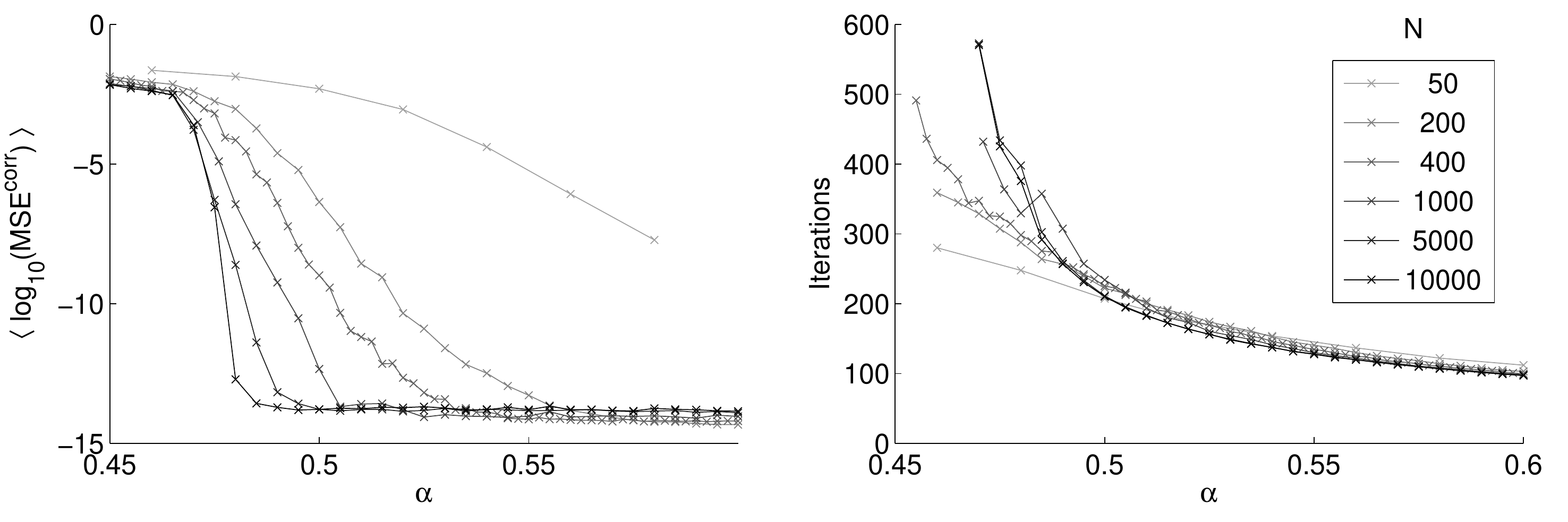}
\caption{Left: C-AMP phase transition as the system
  size $N$ grows. The curves are obtained by averaging $\log_{10}({\rm
    MSE}^{\rm corr})$ over $100$ samples, reflecting the probability of correct
  reconstruction in the region close to the phase transition, where
  it is not guaranteed. Parameters are: $\rho=0.2$, $P=2$,
  $\sigma^2=0.0251$. For higher values of $N$, the phase
  transition becomes sharper.  Right: Mean number of iterations
  necessary for reconstruction, when the true signal {\it is} successfully
  recovered. Far from the phase transition, increasing $N$ does not
  increase visibly the number of iterations for these system sizes,
  showing that our algorithm works in linear time. The number of
  needed iterations increases drastically as one approaches the phase transition. }
\label{Fig:Transition}
\end{figure}

\begin{figure}[!ht]
\centering
\includegraphics[width=0.4\textwidth]{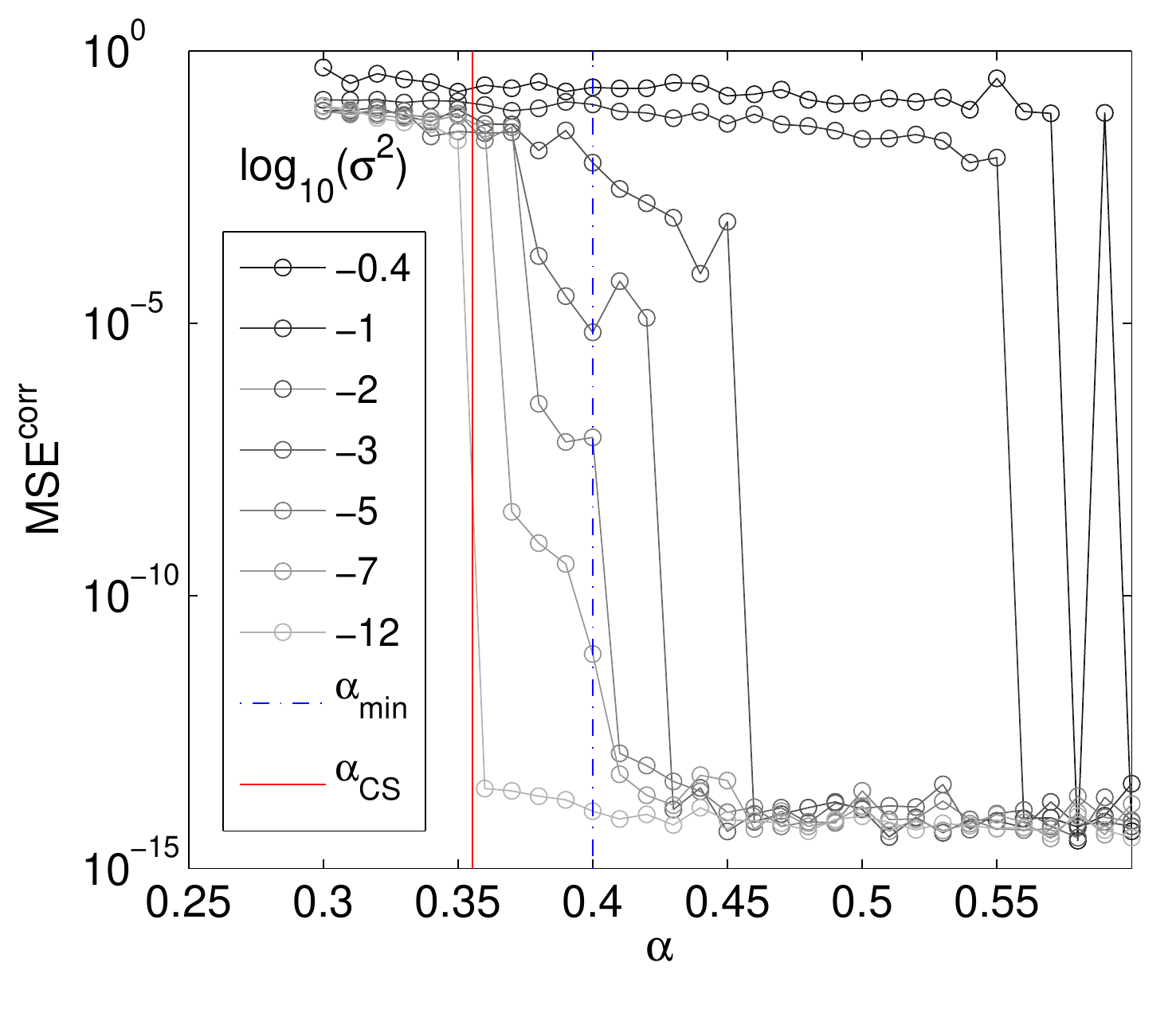}
\includegraphics[width=0.4\textwidth]{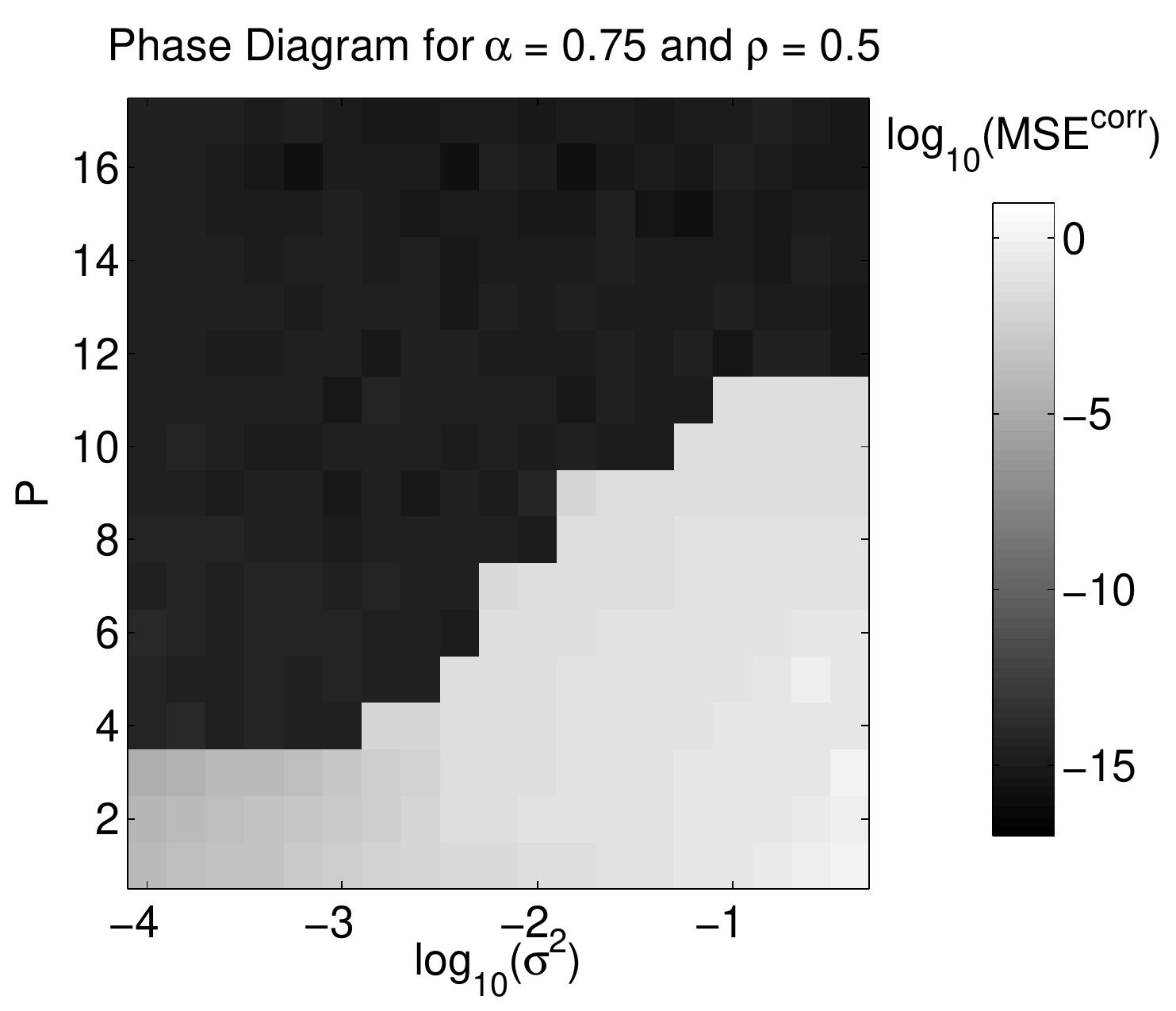}
\caption{Left: Position of the phase transition in $\alpha$ for different 
  distortion variances $\sigma^2$. The left vertical line represents 
the position of the compressed sensing phase transition, 
the right one is the counting bound eq.~(\ref{counting}).
With growing distortion, larger measurement rates become necessary for perfect
calibration and reconstruction.  Intermediary 
values of ${\rm MSE}^{\rm corr}$ are obtained in a region where
perfect calibration is not possible, but 
distortions are small enough for the uncalibrated AMP to make only
small mistakes.
The parameters are those of 
Fig.~\ref{Fig:ARdiags}, and $\alpha$ takes values on the green segment plotted on the $P=2$ diagram.
Right: Phase diagram as the variance of the distortions $\sigma^2$ and
the number of signals $P$ vary, for $\rho=0.5$, $\alpha=0.75$ and $N=1000$. 
}
\label{Fig:SigmaInfluence}
\end{figure}

Fig.~\ref{Fig:Transition} shows the behavior near the phase
transition, giving insights about the influence of the system size and
the number of iterations needed for precise calibration and
reconstruction.  In Fig.~\ref{Fig:SigmaInfluence}, we show the jump in
the MSE on a single instance as the measurement rate $\alpha$
decreases. The right part is the phase diagram in the $\sigma^2$-$P$
plane.

\begin{figure}[!ht]
\begin{center}
\includegraphics[width=1\textwidth]{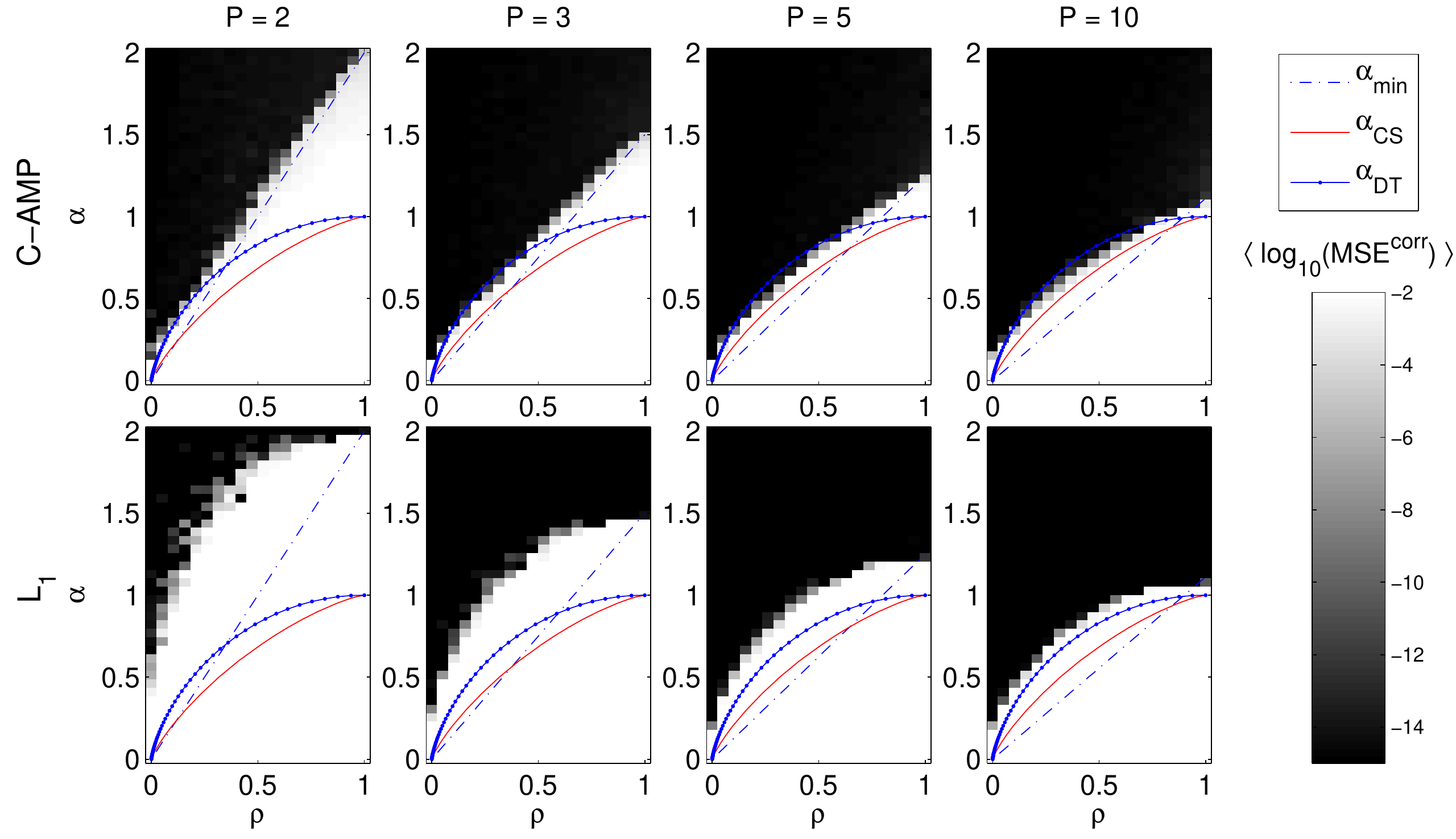}
\end{center}
\caption{Comparison of the empirical phase diagrams obtained with the
  C-AMP algorithm proposed here (top) and the $\ell_1$-minimization
  algorithm of \cite{GribonvalChardon11} (bottom) averaged over
  several random samples; white indicates failure, black indicates
  success. The area where reconstruction is possible is consistently
  much larger for C-AMP than for $\ell_1$-minimization. The plotted
  lines are the phase transitions for the pure compressed sensing
  problem with the AMP algorithm ($\alpha_{CS}$, in red, from
  \cite{KrzakalaMezard12}), and with $\ell_1$-minimization
  (the Donoho-Tanner transition $\alpha_{\rm DT}$, in blue, from \cite{Donoho05072005}).  The line
  $\alpha_{\rm min}$ is the lower counting bound from
  eq.~(\ref{counting}). The advantage of C-AMP over
  $\ell_1$-minimization is clear. Note that in both cases, the region close to the transition is
  blurred due to finite system size, hence a region of grey pixels
  (again, the effect is more prononced for the $\ell_1$ algorithm).}
\label{L1Comp}
\end{figure}

In \cite{GribonvalChardon11,BilenPuy13}, a calibration algorithm using
$\ell_1$-minimization have been proposed.  While in that case, no assumption on the distribution
of the signals and of the the gains is needed, for most practical cases it is expected to be less
performant than the C-AMP if these distributions are known or reasonably
approximated. We implemented the algorithm of
\cite{GribonvalChardon11} with MATLAB using the CVX package
\cite{cvx}.  Due to longer running times, experiments were made using
a smaller system size $N=100$. We also remind at this point that
whereas the C-AMP algorithm works for a generic transfer function
(\ref{definition}), the $\ell_1$-minimization based calibration is
restricted to the transfer functions considered by
\cite{GribonvalChardon11,BilenPuy13}.
Fig.~\ref{L1Comp} shows a comparison of the performances of the two
algorithms in the $\alpha$-$\rho$ phase diagrams. The C-AMP clearly
outperforms the $\ell_1$-minimization in the sense that the region in
which calibration is possible is much larger. 

\section{Conclusion}						
We have presented the C-AMP algorithm for blind calibration in
compressed sensing, a problem where the outputs of the measurements
are distorted by some {\it unknown} gains on the sensors,
eq. (\ref{definition}).  The C-AMP algorithm allows to jointly infer
sparse signals and the distortion parameters of each sensor even with
a very small number of signals and is computationally comparable to
the GAMP algorithm \cite{Rangan10b}.  Another advantage
w.r.t. previous works is that the C-AMP algorithm works for generic
transfer function between the measurements and the readings from the
sensor, not only those that permit a convex formulation of the
inference problem as in \cite{GribonvalChardon11,BilenPuy13}. In the
numerical analysis, we focussed on the case of the product transfer
function (\ref{product}) studied in~\cite{GribonvalChardon11}.  Our
results show that, for the chosen parameters, calibration is possible
with a very small number of different sparse signals $P$
(i.e. $P=2$ or $P=3$), even very close to the absolute minimum of
measurements required by a counting bound (\ref{counting}).
Comparison with the $\ell_1$-minimizing algorithm clearly shows lower
requirements on the measurement rate $\alpha$ and on the number of
signals $P$ for C-AMP. The C-AMP algorithm for blind calibration is
scalable and simple to implement. The efficiency of blind
(unsupervised) calibration for compressed sensing shows that the
knowledge of the training signals is not necessary. We expect C-AMP to
become useful in practical compressed sensing implementations.

Asymptotic analysis of the C-AMP algorithm can be done using the state
evolution approach \cite{DonohoMaleki09}.  In the present case of
blind calibration, however, the resulting equations include a $P$-uple
integration that is numerically demanding and has hence been postponed
to future work.  Future work also includes the study of the robustness
to the mismatch between assumed and true distribution of signal
elements and distortion parameters, as well as the
expectation-maximization based learning of the various parameters.
Finally, the use of spatially coupled measurement matrices
\cite{KrzakalaPRX2012,DonohoJavanmard11} could further improve the
performance of the algorithm and make the phase transition coincide
with the information-theoretical counting bound (\ref{counting}).

\newpage



\bibliographystyle{unsrt}
\bibliography{refs}

\end{document}